\documentclass[nofootinbib,amsfonts,prd,aps]{revtex4}
\usepackage{graphicx}
\begin{document}

\title{Loop quantization of the Schwarzschild black hole}

\author{Rodolfo Gambini$^{1}$,
Jorge Pullin$^{2}$}
\affiliation {
1. Instituto de F\'{\i}sica, Facultad de Ciencias, 
Igu\'a 4225, esq. Mataojo, 11400 Montevideo, Uruguay. \\
2. Department of Physics and Astronomy, Louisiana State University,
Baton Rouge, LA 70803-4001}

\begin{abstract}
  We quantize spherically symmetric vacuum gravity without gauge
  fixing the diffeomorphism constraint. Through a rescaling, we make
  the algebra of Hamiltonian constraints Abelian and therefore the
  constraint algebra is a true Lie algebra. This allows the completion
  of the Dirac quantization procedure using loop quantum gravity
  techniques. We can construct explicitly the exact solutions of the
  physical Hilbert space annihilated by all constraints. New
  observables living in the bulk appear at the quantum level
  (analogous to spin in quantum mechanics) that are not present at the
  classical level and are associated with the discrete nature of the
  spin network states of loop quantum gravity. The resulting quantum
  space-times resolve the singularity present in the classical theory
  inside black holes.  
\end{abstract}

\maketitle

Spherically symmetric gravity in vacuum is perhaps one of the simplest
symmetry reduced models to be studied where there is spatial
dependence of the variables.  In particular, it includes the
interesting case of having a black hole present, with the challenge of
its singularity. There have been previous investigations of the
quantization of vacuum spherically symmetric gravity using complex
Asthekar variables by Thiemann and Kastrup \cite{thiemannsph},
traditional metric variables by Kucha\v{r} \cite{kuchar} and using
modern loop quantum gravity techniques by Campiglia {\em et al.}
\cite{spherical} and Tibrewala \cite{tibrewala}. In all cases the
procedure started by choosing variables adapted to spherical
symmetry. The resulting model has a diffeomorphism constraint
associated with the symmetry under re-scalings of the radial
coordinate and a Hamiltonian constraint representing invariance under
different foliations of space-time. Thiemann and Kastrup
\cite{thiemannsph} were the first to complete the quantization of the
model, remarkably, using Ashtekar's original complex variables. They
noted that essentially there is only one degree of freedom, the
Arnowitt--Deser--Misner (ADM) mass, that does not evolve in
time. Using traditional variables and a suitable set of canonical
transformations, Kucha\v{r} \cite{kuchar} reaches the same result.
Bojowald and
Swiderski \cite{boswi} studied the model in terms of modern, real,
Ashtekar variables and encountered difficulties in performing a canonical
quantization using standard \cite{qsd} loop quantum gravity
techniques.  Based on that work, a loop quantization was achieved by
Campiglia {\em et al.} partially fixing the gauge, which eliminates
the diffeomorphism constraint. Again, the only degree of freedom left
is the ADM mass. Wavefunctions are functions of the ADM mass and if
one reconstructs the metric back from them one has a singularity where
the classical theory had one, the quantization being equivalent to the
one found by Thiemann and Kastrup, and Kucha\v{r}. However, a later
treatment using the semiclassical equations resulting from loop
quantum gravity and covering both the interior and exterior of the
black hole suggested that the singularity could be eliminated
\cite{complete}. Studies of the quantization of black hole interiors
using the isometry with the Kantowski--Sachs space-time also suggested
that the singularity is eliminated by loop quantum gravity \cite{ks}.

In this paper we would like to show that one can proceed to quantize
these models without further gauge fixing. In principle that would be
problematic because the constraint algebra of general relativity, even
in this simple $1+1$ dimensional example, is not a Lie algebra and
that leads to problems implementing the Dirac quantization
procedure. We will show, however, that through a simple rescaling of
the Hamiltonian constraint without changing the canonical variables,
one ends up with a true Lie algebra and can complete the
quantization. In particular one can find exactly the space of physical
states annihilated by the constraints. Using the type of measures
\cite{ashtekarlewandowski} common in loop quantization one can show
that the singularity is eliminated and one ends up with a regular
space-time that tunnels through where the classical singularity used
to be into another universe, akin to what happens classically in the
Reissner-Nordstr\"om space-time, but without singularities. The quantum
theory has additional observables than the ADM mass of the space-time,
related to the fact that at the Planck scale one has structure when
one introduces the types of measures one uses in loop quantum
gravity. These types of degrees of freedom associated to the bulk suggest
that it is possible to have loss of information either via the region
of high curvature that replaces the singularity or through the bulk
observables.

The treatment of spherically symmetric space-times with Ashtekar-type
variables was pioneered by Bengtsson \cite{bengtsson} and in more
modern language discussed in detail by Bojowald and
Swiderski \cite{boswi}. We will follow here the notation of our
previous paper \cite{spherical} and we refer the reader to them and to
Bojowald and Swiderski for more details.

Using Ashtekar-like variables adapted to the symmetry of the problem,
one is left with two pairs of canonical variables $E^\varphi$,
${K}_\varphi$ and $E^x$, $K_x$, that are related to the traditional
canonical variables in spherical symmetry $ds^2=\Lambda^2 dx^2+R^2
d\Omega^2$ by $\Lambda=E^\varphi/\sqrt{|E^x|}$, $P_\Lambda=
-\sqrt{|E^x|}K_\varphi$, $R=\sqrt{|E^x|}$ and $P_R=-2\sqrt{|E^x|} K_x
-E^\varphi K_\varphi/\sqrt{|E^x|}$ where $P_\Lambda, P_R$ are the
momenta canonically conjugate to $\Lambda$ and $R$ respectively, $x$
is the radial coordinate and $d\Omega^2=d\theta^2+\sin^2\theta
d\varphi^2$. We are taking the Immirzi parameter equal to one. For
most of the paper we will analyze the region with $E^x>0$ and we will
therefore drop the absolute value signs inside the square roots, if
one wishes to analyze other regions, the absolute value signs should
be reinstated.

The total Hamiltonian density for the theory is given by 
\begin{equation}
  H_T =\! N\!\!
\left[\frac{\left((E^x)'\right)^2}{8\sqrt{E^x}E^\varphi}
-\frac{E^\varphi}{2\sqrt{E^x}} - 2 K_\varphi \sqrt{E^x} K_x  
-\frac{E^\varphi K_\varphi^2}{2 \sqrt{E^x}}
-\frac{\sqrt{E^x}(E^x)' (E^\varphi)'}{2 (E^\varphi)^2} +
\frac{\sqrt{E^x} (E^x)''}{2 E^\varphi}\right]-N_r
\left[(E^x)' K_x -E^\varphi K_\varphi'\right].
\end{equation}
We proceed to rescale the Lagrange multipliers,

$  N_r^{\rm old}=N_r^{\rm new} -2 N^{\rm
  old}\frac{K_\varphi\sqrt{E^x}}{\left(E^x\right)'}$ and 
  $N^{\rm old} = N^{\rm new} \frac{\left(E^x\right)'}{E^\varphi}$,
and from now on we will drop the ``new'' subscripts. This leads to a
total Hamiltonian that, after an integration by parts, reads,
\begin{equation}
H_T =\int dx \left[ -N'
\left(-\sqrt{E^x}\left(1+K_\varphi^2\right)+\frac{\left(\left(E^x\right)'\right)^2\sqrt{E^x}}{4
    \left(E^\varphi\right)^2}+2 G M\right)
+ N_r \left[-
(E^x)' K_x +E^\varphi K_\varphi'\right]\right] 
\end{equation}
we are not including contributions at the boundary for simplicity. The
constant of integration $2GM$ is obtained imposing the boundary
conditions for the lapse. The discussion in detail is present in
\cite{kuchar}.  A remarkable fact is that this rescaling of the
constraints makes the Hamiltonian constraint have an Abelian algebra
with itself, and the usual algebra with the diffeomorphism
constraint. We had already noted this in \cite{spherical} but after
gauge fixing the diffeomorphism constraint, here we point out that it
is true even without gauge fixing the diffeomorphism constraint.

We now proceed to quantize. We start by recalling   the basis of
spin network states in one dimension (see \cite{spherical} for
details). One has  graphs $g$ consisting of a collection of edges
$e_j$ connecting
the vertices $v_j$. 
It is natural to associate the variable $K_x$ with
edges in the graph and the variable $K_\varphi$ with vertices of the
graph. For bookkeeping purposes we will associate each edge with the
vertex to its left. One then constructs a standard holonomy for $K_x$
and a ``point holonomy'' for $K_\varphi$ (since it behaves as a scalar),
\begin{equation}
T_{g,\vec{k},\vec{\mu}}(K_x,K_\varphi) = \langle K_x,K_\varphi
\left\vert
g,\vec{k},\vec{\mu}
\right\rangle=
\prod_{e_j\in g}
\exp\left(\frac{i}{2} k_{j} \int_{e_j} K_x(x)dx\right)
\prod_{v_j\in g}
\exp\left(\frac{i}{2}  \mu_{j} K_\varphi(v_j) \right)
\end{equation}
with $e_j$ the edges of the spin network $g$ and $v_j$ its vertices
and the integer $k_j$ is the ``color'' associated with the edge $e_j$
and the real number $\mu_j$
the ``color'' associated with the vertex $v_j$.
On these states the triads act multiplicatively,
\begin{eqnarray}
{\hat{E}^x(x) } T_{g,\vec{k},\vec{\mu}}(K_x,K_\varphi)
&=& \ell_{\rm Planck}^2 k_i(x) T_{g,\vec{k},\vec{\mu}}(K_x,K_\varphi),
\\
\hat{E}^\varphi(x) T_{g,\vec{k},\vec{\mu}}(K_x,K_\varphi)
&=& \ell_{\rm Planck}^2 \sum_{v_i\in g} \delta(x-x(v_i))\mu_i 
T_{g,\vec{k},\vec{\mu}}(K_x,K_\varphi),
\end{eqnarray}
where $k_i(x)$ is the color of the edge including the point $x$. If
the
latter is at a vertex, it is the edge to the right of it.
$x(v_i)$ is the position of the vertex $v_i$.

To deal with the Hamiltonian constraint we follow the steps usual in
loop quantum cosmology and replace $K_\varphi\to\sin\left(\rho
  K_\varphi\right)/\rho$ in order to have a well defined operator on
the kinematical Hilbert space (some authors choose $\rho=1$
\cite{qsd}). We also choose a factor ordering, and it is convenient to
rescale $H$, and to take a square root to simplify solving it,
\begin{equation}
  \hat{H}(N)=\int dx N(x)\left(  2\left\{\sqrt{\sqrt{\hat{E}^x}
\left(1 +{\sin^2\left(\rho
          \hat{K}_{\varphi}\right)}/{\rho^2}\right) 
-2 G M}\right\}\hat{E}^\varphi -\sqrt[4]{\hat{E}^x}\left(\hat{E}^x\right)'\right)
\end{equation}
and the quantum constraint is also Abelian free of anomalies. 
We have defined the action of the relevant operators involved in the
constraint on the $T_{g,\vec{k},\vec{\mu}}$ basis. Let us recall that
$K_{\varphi}$ is not well defined as an operator, only its
exponentiation, so we had to polymerize, with $\rho$ the
polymerization parameter. Acting on a quantum state we have that,
\begin{eqnarray}
  \hat{H}(N) T_{g,\vec{k},\vec{\mu}}(K_x,K_\varphi) &=&\sum_{v_i\in g}
  N(v_i)\left(k_i \ell_{\rm Planck}^2\right)^{\frac{1}{4}}
\Big[2 \sqrt{1+\frac{\sin^2\left(\rho
          K_{\varphi}(v_i)\right)}{\rho^2}
 -\frac{2GM}{\sqrt{k_i
        \ell_{\rm Planck}^2}}}\ell_{\rm Planck}^2 \mu_i\nonumber\\
&&-  \left(k_i-k_{i-1}\right)\ell_{\rm Planck}^2\Big]
T_{g,\vec{k},\vec{\mu}}(K_x,K_\varphi).
\end{eqnarray}

Seeking, for simplicity, a solution of the form, 
 $\Psi\left(K_{\varphi},K_{x},g,
  \vec{k},M\right) = \sum_{v\in g} \sum_{\mu(v)}
T_{g,\vec{k},\vec{\mu}}(K_x,K_\varphi) \Psi(\mu(v),M),$ (one could
also consider superpositions in $\vec{k}$), the equation
$\hat{H}(N)\Psi=0$ can be solved and leads to,
\begin{equation}
  \Psi\left(K_{\varphi},K_{x},g,\vec{k},M\right) =
  \exp\left(f\left(K_{\varphi},g,\vec{k},M\right)\right) \Pi_{e_j\in g} \exp\left(\frac{i}{2}
  k_j\int_{e_j} K_{x}(x)dx\right),
\end{equation}
with $f$ given by $ f=\sum_{v_j\in g} -\frac{i}{2} \Delta K_j m_j
F\left(\sin\left(\rho K_{\varphi}(v_j), i m_j\right)\right), $ with
$\Delta K_j = K_{\varphi}(v_j)-K_{\varphi}(v_{j-1})$, $m_j=\left[\rho
  \sqrt{1-2 G M/\sqrt{k_{j}}\ell_{\rm Planck}}\right]^{-1}$ and $
F(\phi,m)=\int_0^\phi\left(1-m^2 \sin^2 t\right)^{-1/2} dt$ the Jacobi
(sometimes also known as Legendre) elliptic function of the first
kind.  Although $m_j$ is purely imaginary inside the horizon, one can
show that $\exp(f)$ is a pure phase factor for any value of $k_j$, no
matter if it corresponds to the black hole interior or not.

The above solution of the Hamiltonian constraint is not invariant
under diffeomorphisms. That can be readily corrected via standard group
averaging in which one superposes a family of states related by
diffeomorphisms \cite{ga}. For reasons the space we do not show it explicitly since
the construction is standard. One ends up with states that are
superpositions of spin networks with vertices in all possible
positions along the radial line, preserving the order of them (this
last point will play a crucial role in the appearance of new quantum
observables). In higher dimensions such order could be associated with
the diffeo invariant non-trivial knotting of the spin networks. The
resulting state is a functional of $K_x,K_\varphi$ labeled by a
diffeomorphism-related class of graphs $\tilde{g}$ and the colors
for each edge $\vec{k}$ and the ADM mass $M$. We denote them as $\vert
\vec{k},\tilde{g}\rangle$, omitting the dependence on $M$ for
simplicity. These vectors define a basis for the physical space of
states ${\cal H}_{\rm phys}$. 

The graph $g$ is based on an integer number $V$ of vertices located at
$x(v_1),\ldots x(v_V)$. Since the elements of the basis of the states
of ${\cal H}_{\rm phys}$ have a well defined number of vertices, one
can construct an Dirac observable $\hat{V}$ that acting on $\vert
\vec{k},\tilde{g}\rangle$ has as eigenvalue the integer number $V$.
This is an observable that has no classical counterpart.

Even more interesting is the observable associated with the sequence
of monotonically growing integers $\vec{k}$ that characterizes the
sequence of characteristic radii of black holes. The Hamiltonian
constraint does not change the values of $\vec{k}$ and neither does
the diffeomorphism constraint.  In the classical theory the radial
coordinate $E^x$ is diffeomorphic to $x^2$, but since it can only be a
monotonically growing function this restricts the types of
transformations allowed. To yield a non-monotonic function one would
have to consider diffeomorphisms that are not invertible. That
restriction is what in the quantum theory ends up yielding a new
observable. If one were in more than one dimension, there are similar,
more complex, restrictions arising from the knotting of spin networks.

An operator associated with the sequence $\vec{k}$ that is a
Dirac observable acting on the physical space of states is  $O(z)$ with
$z\in [0,1]$,
$  \hat{O}(z)\vert \vec{k},\tilde{g}\rangle_{\rm phys}= \ell_{\rm
    Planck}^2 k_{{\rm Int}(V z)} 
\vert  \vec{k},\tilde{g}\rangle_{\rm phys}$, 
where ${\rm Int}(V z)$ is the integer part of $V z$ and $V$ is the
number of vertices.  These quantum operators characterize the quantum
geometry and as we shall see may have profound physical implications.
On the physical Hilbert space ${\cal H}_{\rm phys}$, $M$ is a Dirac
observable but $E^x(x)$ is not. However, given an arbitrary monotonic
function from the interval of the radial direction we are studying
(for instance the origin and an asymptotic boundary $[0,x_+]$) to the
interval $[0,1]$, which we call $z(x)$, one has that $\hat{E}^x(x)
\vert \vec{k},\tilde{g}\rangle_{\rm phys} = \hat{O}(z(x))\vert
\vec{k},\tilde{g}\rangle_{\rm phys} $. The function $z(x)$
characterizes the gauge freedom in $E^x$. Recall that the eigenvalues
of $\hat{E}^x$ in the kinematical Hilbert space can only take the
values $\ell_{\rm Planck}^2k_i$. Another way of understanding this
is that we are defining an evolving constant of the motion associated
with $E^x$ that is a function of a ``parameter'' given by the function
$z(x)$ and the observable $\hat{O}(z)$.

In a similar fashion, one can define
evolving constants of the motion that represent the metric of
space-time acting on ${\cal H}_{\rm phys}$. The parameters of the
evolving constant are $K_\varphi$ subject to suitable boundary
conditions and $z(x)$. For instance the classical expression of the
$g_{tx}$ component of the metric is, 
\begin{equation}
 g_{tx} = g_{xx} N_r =-\frac{\left(E^x\right)' K_\varphi}{2
   \sqrt{E^x}\sqrt{1+K_\varphi^2-\frac{2 G M}{\sqrt{E^x}}}},
\end{equation}
with similar expressions for $g_{tt}$ and $g_{xx}$. This can be
derived choosing a gauge in which $K_\varphi$ and $E^x$ are given
functions of space and preserving the gauge fixing conditions in time
through the determination of the lapse and the shift.  For instance,
this would lead to the usual form of the Schwarzchild metric provided
one chooses $K_\varphi^0=0$ and $E^x_0=x^2$. We can proceed to promote
it to a quantum operator on $H_{\rm phys}$ in terms of $\hat{O}(z)$,
$\hat{M}$ and parameterized by $K_\varphi$ and $z(x)$ (the expression
needs to be made well defined first by introducing holonomies),
\begin{equation}
 \hat{g}_{tx} = -\frac{\left(\hat{E}^x\right)' \sin\left(\rho K_\varphi\right)}{2\rho
   \sqrt{\hat{E}^x}\sqrt{1+\frac{\sin^2\left(K_\varphi\right)}{\rho^2}-\frac{2 G \hat{M}}{\sqrt{\hat{E}^x}}}},
\end{equation}

The square root that appears in $\hat{g}_{tx}$ leads 
to the following
inequality, in order to get a self-adjoint operator (notice that there
are no factor ordering issues), 
$
1 +\left(\frac{\sin \rho K_{\varphi}}{\rho}\right)^2-\frac{2 G
  M}{\sqrt{E^x/\epsilon}}\ge 0.
$ The most unfavorable point is when the eigenvalues of $E^x$ become small.
The most favorable gauge choice, from the point of view of keeping the
expression positive at that point is $K_{\varphi,
  0}=\pi/(2\rho)$ where $K_{\varphi,0}$ is $K_\varphi$ evaluated at
$x=0$ (since $E^x$ is monotonic the worse case happens at $x=0$). 
Therefore the gauge
condition for the square root that appears in the metric to be real
and therefore the metric self-adjoint is, in terms of the eigenvalues
of $\hat{E}^x$, given by, 
$
  k_0> \left(\frac{2 G M }{\ell_{\rm Planck}\left(1
        +\frac{1}{\rho^2}\right)}\right)^2.
$ As a consequence, small values of $k_0$  are 
excluded in order to have a self adjoint metric operator and as a
consequence the singularity is avoided.  The region exterior to the
horizon is covered for any gauge $K_\varphi$ since the last term in
the first inequality is less or equal to one outside the
horizon. Notice that there exist gauge choices that would make the
metric singular. Those correspond to coordinate singularities and loop
quantum gravity correctly does not eliminate them.

Although these general considerations about the geometry are true for
any state, obviously not all states lead to semiclassical
geometries. To begin with, the operators corresponding to the metric
are distributional, having support only at vertices of the spin
net. One can envision them approximating smooth geometries for spin
nets with densely packed vertices and some additional conditions: a)
One would need to consider a superposition of ADM masses with a weight
$\Psi(M)$ that is a peaked functions around a given value $M_0$; b)
One also needs to require some smoothness in the radial coordinate by
limiting the jumps of the eigenvalues from a vertex to the next, i.e.
$\Delta k_i<\Delta_0$ with $\Delta_0$ controlling the level of
smoothness. One could in particular, approximate a smooth
Schwarzschild geometry with quantum corrections. Also notice that for
simplicity we have also kept the discussion in terms of states that
are eigenstates of $\hat{O}(z)$. In reality one will have states that
will involve superpositions of values of $\vec{k}$ as well.

The analysis can be extended to the interval $[-x_+,x_+]$ with a
trivial extension of $O(z)$ to $z\in[-1,1]$. The expectation value of
the determinant of the space-time metric can be explicitly calculated
in a suitable gauge and it goes through a maximum value and starts
decreasing for negative values of $x$. One can view this as a
generalization of the Kruskal extension including a new region that
tunnels through the singularity.

We have performed a loop quantization of the vacuum spherically
symmetric space-times. Apart from using variables adapted to spherical
symmetry, we did not perform any additional gauge fixing. Through a
rescaling of the Hamiltonian constraint, the constraint algebra was
turned into a Lie algebra. We were able to exactly solve the
constraints and find the space of physical states. We encounter that
in addition to the ADM mass and its canonically conjugate momentum
other Dirac observables arise in the quantum theory associated with
the bulk of the space-time. The metric of the space-time can be
analyzed as an operator in the physical space of state viewing it as an
evolving constant of the motion, written in terms of the Dirac
observables and free parameters that represent the coordinate
freedom. One sees that the singularity that arises in the classical
theory is eliminated and is replaced by a region of high curvature
through which the space-time could be extended, yielding a global
structure similar to that of the Reissner--Nordstrom space-time but
without singularities, as had been anticipated in a previous
treatment using the effective semi-classical theory \cite{complete}.

The existence of the new quantum observables, and the associated
degrees of freedom, may have some relevance
for the recent discussion of ``firewalls'' in black hole evaporation.
Almheiri {\em et al.}  \cite{amps} (and earlier
Braunstein {\em et al.}  \cite{braunstein}) showed that in
order to preserve the unitarity of the $S$ matrix during black hole
evaporation drastic changes in the usual picture were needed, like
surrounding the black hole with a firewall. This follows from
fundamental hypotheses, like the existence of a unitary $S$ matrix
that describes the evolution of the incoming pure state that forms the
black hole and the outgoing Hawking radiation. From the perspective of
our analysis, this hypothesis is not obvious since in principle there
could be part of the information lost when falling into the black hole
interior tunneling into another region or into the new local degrees
of freedom we discussed. Our analysis is at the moment
limited to the vacuum case. However, from the form of the Hamiltonian
constraint coupled to matter one can see that the bulk observables
persist in that case, suggesting that the analysis of the information
issue made could be carried out explicitly in the case of an
evaporating black hole.

We wish to thank Jerzy Lewandowski, Don Marolf and Javier Olmedo for comments.
This work was supported in part by grant NSF-PHY-0968871, funds of the
Hearne Institute for Theoretical Physics, CCT-LSU and Pedeciba.
%\vspace{-7mm}

\end{document}